\documentclass[aps,prl,twocolumn,superscriptaddress,reprint,showpacs]{revtex4-1}
\usepackage{graphicx}
\usepackage{setspace}
\usepackage{amsmath}
\usepackage{amssymb}
\usepackage{color}
\usepackage{array}
\usepackage{subfigure}
\usepackage{hyperref}
\usepackage{float}
\usepackage{lipsum}

\usepackage[all]{xy}
\newcommand{\RN}[1]{%
  \textup{\uppercase\expandafter{\romannumeral#1}}%
}

\begin{document}
\title{Nearly nondestructive thermometry of labeled cold atoms and application to isotropic laser cooling}

\author{Xin Wang}
\affiliation{Key Laboratory of Quantum Optics and Center of Cold Atom Physics, Shanghai Institute of Optics and Fine Mechanics, Chinese Academy of Sciences, Shanghai 201800, China}
\affiliation{Center of Materials Science and Optoelectronics Engineering, University of Chinese Academy of Sciences, Beijing 100049, China}
\author{Yuan Sun}
\email{yuansun@siom.ac.cn}
\affiliation{Key Laboratory of Quantum Optics and Center of Cold Atom Physics, Shanghai Institute of Optics and Fine Mechanics, Chinese Academy of Sciences, Shanghai 201800, China}
\author{Hua-Dong Cheng}
\affiliation{Key Laboratory of Quantum Optics and Center of Cold Atom Physics, Shanghai Institute of Optics and Fine Mechanics, Chinese Academy of Sciences, Shanghai 201800, China}
\affiliation{Center of Materials Science and Optoelectronics Engineering, University of Chinese Academy of Sciences, Beijing 100049, China}
\author{Jin-Yin Wan}
\affiliation{Key Laboratory of Quantum Optics and Center of Cold Atom Physics, Shanghai Institute of Optics and Fine Mechanics, Chinese Academy of Sciences, Shanghai 201800, China}
\author{Yan-Ling Meng}
\affiliation{Key Laboratory of Quantum Optics and Center of Cold Atom Physics, Shanghai Institute of Optics and Fine Mechanics, Chinese Academy of Sciences, Shanghai 201800, China}
\author{Ling Xiao}
\affiliation{Key Laboratory of Quantum Optics and Center of Cold Atom Physics, Shanghai Institute of Optics and Fine Mechanics, Chinese Academy of Sciences, Shanghai 201800, China}
\author{Liang Liu}
\email{liang.liu@siom.ac.cn}
\affiliation{Key Laboratory of Quantum Optics and Center of Cold Atom Physics, Shanghai Institute of Optics and Fine Mechanics, Chinese Academy of Sciences, Shanghai 201800, China}

\begin{abstract} 
We have designed and implemented a straightforward method to deterministically measure the temperature of the selected segment of a cold atom ensemble, and we have also developed an upgrade in the form of nondestructive thermometry. The essence is to monitor the thermal expansion of the targeted cold atoms after labeling them through manipulating the internal states, and the nondestructive property relies upon the nearly lossless detection via driving a cycling transition. For cold atoms subject to isotropic laser cooling, this method has the unique capability of addressing only the atoms on the optical detection axis within the enclosure, which is exactly the part we care about in major applications such as atomic clock or quantum sensing. Furthermore, our results confirm the sub-Doppler cooling features in isotropic laser cooling, and we have investigated the relevant cooling properties. Meanwhile, we have applied the recently developed optical configuration with the cooling laser injection in the form of hollow beams, which helps to enhance the cooling performance and accumulate more cold atoms in the central regions. 
\end{abstract}
\maketitle

Ever since the advent of the laser cooling and trapping \cite{PhysRevLett.48.596, PhysRevLett.55.48}, cold atoms have become a key platform in emerging quantum technologies, including quantum precision measurement \cite{RevModPhys.83.331}, quantum sensing \cite{RevModPhys.89.035002}, quantum simulation \cite{RevModPhys.83.1523, nature14223, nature24622} and quantum information \cite{RevModPhys.82.1041, RevModPhys.82.2313}. Over the last three decades, intense efforts have been devoted to the development of relevant technologies, where deterministic measurement of cold atoms' temperature is an essential subject \cite{PhysRevLett.61.169, PhysRevLett.61.826}. So far, time-of-flight (TOF) measurement of falling atoms has become the standard technique to evaluate the temperature under various scenarios such as optical molasses, magneto-optical trapping (MOT) and BEC \cite{PhysRevLett.96.130404, PhysRevLett.103.223203}. Nevertheless, recent progress raises demands for different types of temperature measurement techniques, especially for the situations not so favorable for TOF method. For example, a lot of interest has been attracted to the thermometry of cold atoms in experiments of cavity quantum electrodynamics \cite{PhysRevA.87.033832}, ultracold atoms on the nK scale \cite{PhysRevA.82.011611, PhysRevA.96.062103, nature41586Becker} or subject to strong interactions \cite{PhysRevA.97.063619, PhysRevB.98.045101}, and many other interesting cases in cold atom physics \cite{McKay_2010, Gring1318, PhysRevA.93.043607}. Moreover, nondestructive thermometry of cold atoms \cite{PhysRevA.75.033803} also draws a lot of attention, and a recent study further demonstrates the possibility of the nondestructive temperature measurement technique for BEC \cite{PhysRevLett.122.030403}.

Meanwhile, isotropic laser cooling (ILC) in the form of cold atoms enclosed in an environment of diffuse reflection light \cite{PhysRevLett.69.2483, PhysRevA.49.2780, Wang1995, Guillot:01, PhysRevA.79.023407} has found critical applications in microwave atomic clocks, spectroscopic studies and quantum sensors \cite{Zhang:09_OE, PhysRevA.82.033436, PhysRevA.92.062101, Liu:16_JOSAb, PhysRevApplied.10.064007, PhysRevA.97.023421} due to its unique characteristics of compactness and robustness \cite{PhysRevA.92.062101, PhysRevApplied.10.064007}. To ensure the quality of the diffuse optical field of ILC, the enclosure of the diffuse reflection surface around the atoms needs to cover nearly the entire $4\pi$ solid angle. And therefore, unlike the typical case of six-beam optical molasses or MOT, if we want to apply TOF method to ILC \cite{Guillot:01, PhysRevA.79.023407}, an extra falling distance and an auxiliary chamber equipped with detection windows are necessary, which bears considerable difficulties against the natural requirement of ILC platform. On the other hand, it is also practically troublesome to incorporate a transparent window to image the entire ensemble, such that fluorescent light does not likely become an ideal option for the signal of temperature measurement either. Therefore, it remains as a challenge to find a method that is not only accurate in result but also straightforward in implementation and integration, for temperature assessment of ILC. 

In this letter, we report our recent progress in realizing and characterizing a different method to directly measure the temperature of a selected segment of cold atoms via detecting their thermal expansion after labeling them. The expansion process is monitored based upon the absorption of a probe laser, and the collective information of the density change in time is extracted without the necessity of imaging. First, we present its basic principles and typical procedures of operation. Then, we move on to discuss an upgrade of this method aiming at nearly nondestructive detection. Finally, according to the result of temperature assessment, we investigate the elementary properties of ILC's sub-Doppler cooling effects. This method is virtually applicable to most types of cold atom platforms, and here we put emphasis on its application to ILC experimentally.

\begin{figure}[t]
\centering
\includegraphics[width=0.48\textwidth]{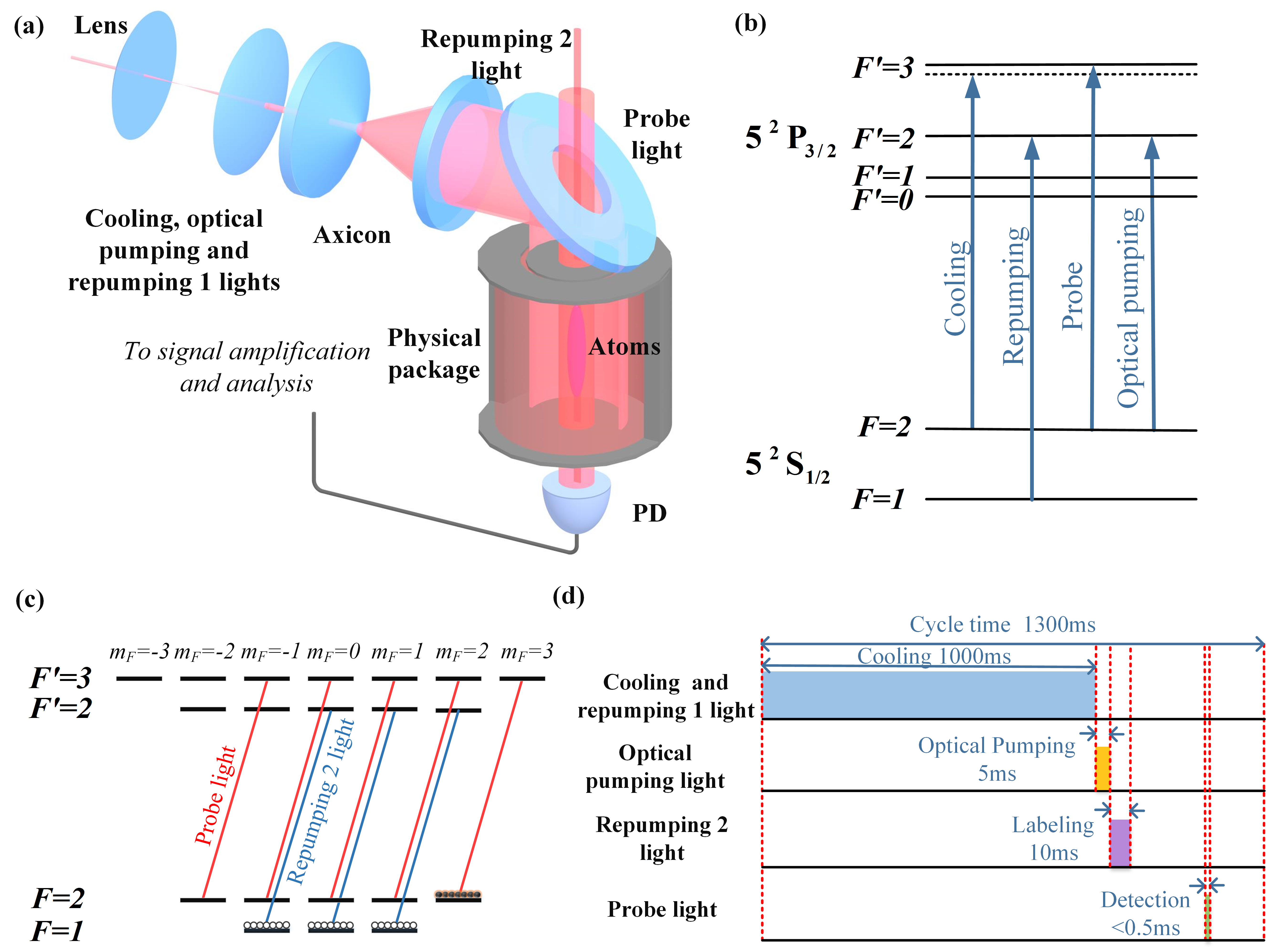}
\caption{(a) Schematic of the experimental setup. The cylindrical cavity is mounted vertically and the cold atom ensemble is of the long cigar shape. The optical detection axis is along the central axis of the cylindrical cavity. The probe light propagates through the cold atoms and arrives at a photodetector (PD). (b) Relevant energy levels and transitions of $^{87}$Rb D2 line for our experiment. (c) Details of the Zeeman sub-states and relevant polarization degrees of freedom for the nearly nondestructive detection method. (d) A sample time sequence. Firstly, atoms are cooled down via ILC. Secondly, optical pumping light transfers all intracavity cold atoms to the level of $5^{2}$S$_{1/2}$, $F=1$. Afterwards, repumping 2 light populates the atoms to $5^{2}$S$_{1/2}$, $F=2$ along the central axis. Finally, during the stage of detection, probe light is applied with a variable delay time which can be scanned.}
\label{fig:experimental_configuration}
\end{figure}

The basic mechanism and implementation are outlined in Fig. \ref{fig:experimental_configuration}. In the experiment, $^{87}$Rb atoms are laser cooled by the diffuse reflection light inside a cylindrical cavity \cite{SuppInfo, Meng_2013_CPL, PhysRevA.97.023421}. The cylindrical cavity is made of a glass cell with inner diameter of 54 mm and height of 54 mm, and its surface is coated with the reflective material whose diffuse reflection index is more than 98\% at the wavelength of 780 nm. The cooling light is red detuned at 22.5 MHz to the transition of $5^{2}\text{S}_{1/2}, F=2 \leftrightarrow 5^{2}\text{P}_{3/2}, F=3$. The repumping light, tuned close to the resonance of $5^{2}\text{S}_{1/2}, F=1 \leftrightarrow 5^{2}\text{P}_{3/2}, F=2$, is divided into two beams. One beam (repumping 1) participates the laser cooling process and the other beam (repumping 2) enforces the labeling process of the atoms along the central axis. The cooling, optical pumping and repumping 1 lights combine into one beam, and then form a parallel hollow beam with diameter of 34 mm and width of 1 mm via a pair of axicons. Eventually, the hollow beam is injected into the cylindrical cavity and gets diffusely reflected inside the cavity. The repumping 2 and probe lights propagate vertically along the central axis through the cavity, with Gaussian beam diameters of 0.96 mm and 2.63 mm respectively. 

Our labeling process is accomplished by distinguishing the two hyperfine ground levels. After interaction with optical pumping light and repumping 2 light, the cold atoms along central axis are pumped to $F=2$ while the rest of the intracavity cold atoms stay in $F=1$. In such an arrangement, it does not involve the polarization degree of freedoms. Here, the target segment is naturally chosen as the cold atoms along the central axis, exactly those we care about in typical applications of ILC platform. The labeled atoms will subsequently start the diffusion process, namely expanding freely due to their thermal motions. Monitoring the diffusion process over time will yield the information of temperature.

\begin{figure}[b!]
\centering
\includegraphics[width=0.48\textwidth]{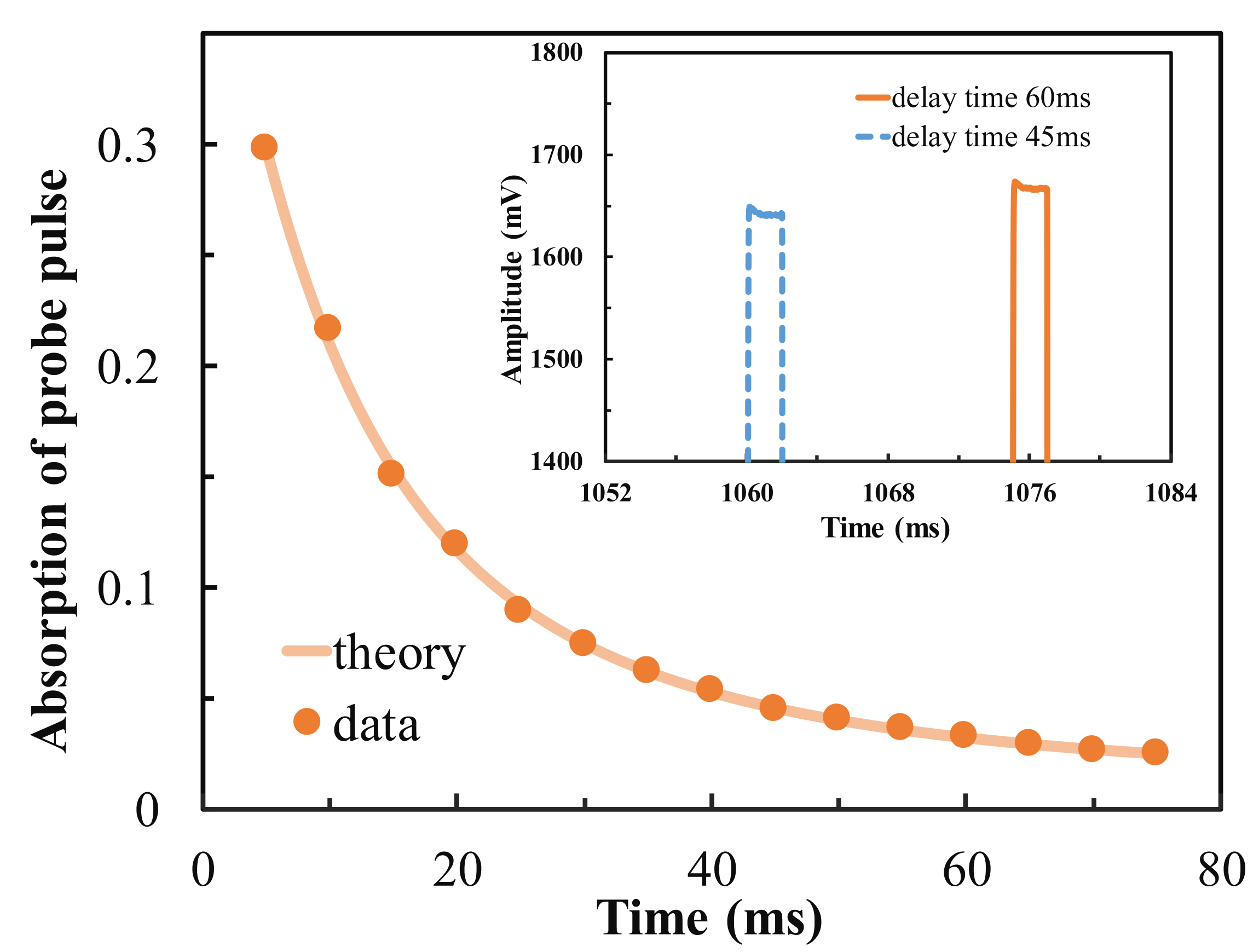}
\caption{A typical result with each data point taken in a new cycle of experiment. The cooling laser power is 120 mW and the cooling time is 1000 ms. The inset shows the raw data of amplified voltage signals from probe pulse incidences on the PD, time stamped with respect to a standard experimental cycle. The error bar is smaller than the data dot size and therefore omitted for now. The theoretical curve according to Eq. \eqref{eq:probe_reduction_temperature} is shown, and the temperature is deduced to be 41.9 $\pm$ 2.6 $\mu$K with the factor of $3/2$ already included.}
\label{fig:sample_experimental_data}
\end{figure}

Fig. \ref{fig:sample_experimental_data} shows the outcome of such a measurement. For simplicity, we neglect the finite width of the labeled atom ensemble, and effectively treat it as a thin line \cite{SuppInfo}. Then after diffusion, the labeled atoms' spatial density profile at a later time directly reflects their original velocity distribution at the beginning. Because the time duration of diffusion is relatively short and the direction of gravity is along the central axis, the effects of gravity can also be virtually dismissed for now. With respect to the geometry of this system, dynamics in the dimension along the central axis does not have a significant impact on the outcome, and therefore it suffices to consider the process under two-dimensional setting in the transverse plane. In laser cooling, assigning temperature to a cold atom ensemble requires careful clarification, since the ensemble does not necessarily sit in a thermal equilibrium state. For our case, the justification comes from the information of the velocity distribution profile, which can be deduced from analyzing the experimental data. It turns out to be well described by the 2D Maxwell-Boltzmann distribution $f(v)dv=\big(\frac{m}{2\pi k_{B}T}\big)2 \pi v \exp\big(-\frac{mv^{2}}{2k_{B}T}\big)dv$. According to the equipartition theorem, the temperature in 3D can be deduced as $T_\text{3D} = T_\text{2D}\times3/2$, by the inherent isotropic property of ILC in the kinematics of each dimension.

Loosely speaking, our method behaves more or less equivalently as counting the number of labeled atoms within a prescribed confined region. Hotter samples tend to diffuse faster with fewer residual atoms, and vice versa. After time delay $t$, the number of atoms remaining in the region with radius $r_{c}$ is proportional to \cite{SuppInfo}:
\begin{equation}
\label{eq:simple_atom_counting}
s(r_{c},t)=1-\exp(-\frac{m(r_{c}/t)^{2}}{2k_{B}T}).
\end{equation}
Scattering off these atoms will lead to a reduction in the transmission of the probe laser pulse. Under the assumption that the cold atomic gas is relatively dilute and the delay time is not too short, the reduction of probe pulse power is proportional to \cite{SuppInfo}:
\begin{equation}
\label{eq:probe_reduction_temperature}
\Big( \frac{2}{(d/2t)^2} + \frac{m}{2k_B T} \Big)^{-1},
\end{equation}
where $d$ is the Gaussian diameter of the probe light.

\begin{figure}[bh]
\centering
\includegraphics[width=0.48\textwidth]{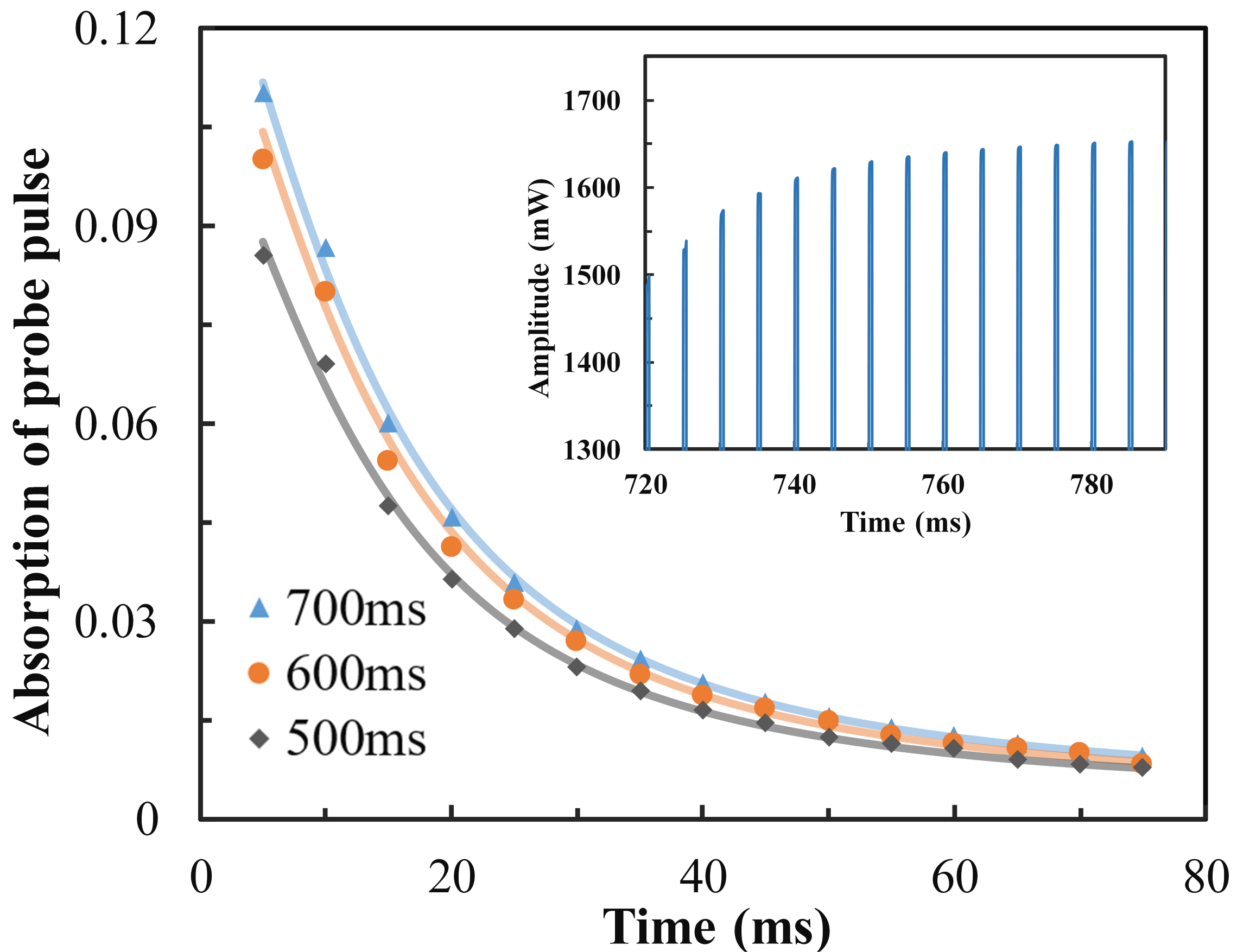}
\caption{Result of temperature measurement by the nearly nondestructive type method, for cold atoms with different cooling times under ILC. According to curve fitting via Eq. \eqref{eq:probe_reduction_temperature}, the temperature difference of these three cases is within $\sim 0.5$ $\mu$K. The differences in absorption between these curves are mostly due to the differences of the accumulated cold atom numbers, which depend on the cooling time. All the cooling laser power is set at 60 mW. The inset shows a sequence of raw data in terms of amplified probe light signal, all taken consecutively in the same one experiment cycle.}
\label{fig:sample_nondestructive}
\end{figure}

Typical cold atom temperature measurement techniques such as TOF or imaging after release are destructive type detections. On the other hand, evaluation of temperature via nondestructive measurement has been of essential interest in the field, and such methods have already been discussed \cite{PhysRevA.75.033803}. The basic experiment of Fig. \ref{fig:sample_experimental_data} is also destructive in the sense that interaction with the probe pulse changes the internal quantum state of cold atoms and the original labeling does not persist. Hence, a complete measurement trace requires repeated trials of replenished cold atoms. However, if we include the polarization degrees of freedom into the system, nearly nondestructive type method can be constructed with the help of the cycling transition, just like the case of atomic qubit readout \cite{PhysRevLett.106.133002, PhysRevLett.106.133003, PhysRevA.92.042710, PhysRevLett.119.180503, PhysRevLett.119.180504}. From the viewpoint of absorption imaging of cold atoms \cite{Gajdacz:13, Wigley:16}, we note that our detection process has connections with optical imaging via compressive sampling techniques \cite{IEEE4472240, Katz2009APL, Han2012APL, Li2019Optica}.

More specifically, for $^{87}$Rb, an appropriate cycling transition can be chosen as $5\text{S}_{1/2} |F=2, m_F=2\rangle \leftrightarrow 5\text{P}_{3/2} |F=3, m_F=3\rangle$, as shown in Fig. \ref{fig:experimental_configuration}(c). Accordingly, we introduce necessary changes into the labeling process such that the labeled atoms are sent to $|F=2, m_F=2\rangle$ while the rest atoms stay in level of $F=1$ \cite{SuppInfo}. Then, interaction of a weak probe pulse of right circular polarization won't destroy the prescribed labeling and the nearly nondestructive detection of temperature can be realized. Magnetic bias field is not necessary for this. In particular, it does not require complicated interferometer design or delicate imaging setup. 

\begin{figure}[b]
	\centering
	\includegraphics[width=0.48\textwidth]{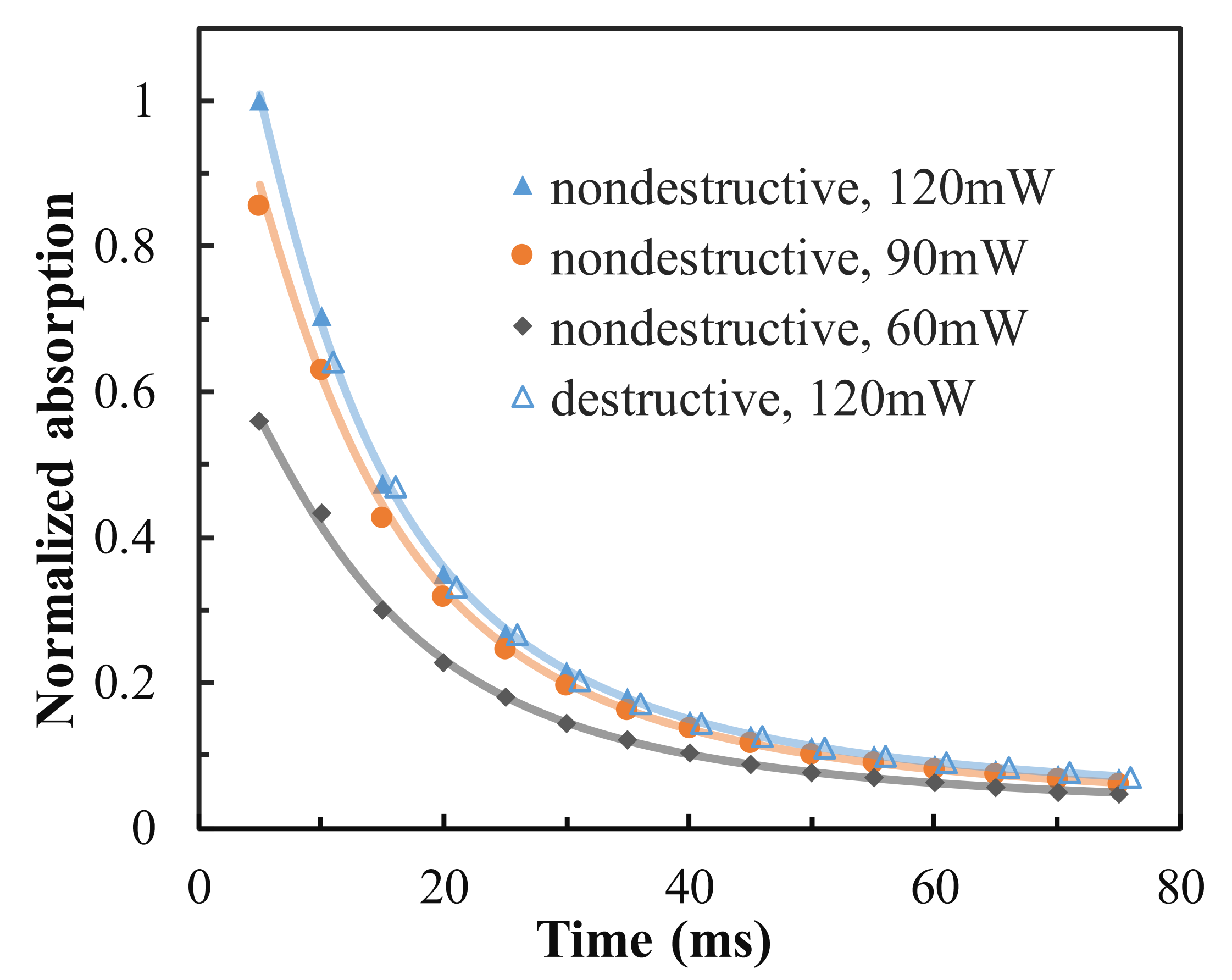}
	\caption{Temperature measurement with respect to different cooling laser powers via the nearly nondestructive measurement, plotted in terms of normalized absorption of probe pulse at different delay times. A comparison between destructive and nondestructive type methods is also included. All nondestructive type measurement data points are normalized with the same standard while the destructive type counterpart is normalized differently for better visualization of the comparison. All the cooling time is set as 1000 ms.}
	\label{fig:nondestructive_compare}
\end{figure}

\begin{figure}[h]
\centering
\begin{tabular}{l}
\includegraphics[trim = 0mm 0mm 0mm 0mm, clip, width=0.48\textwidth]{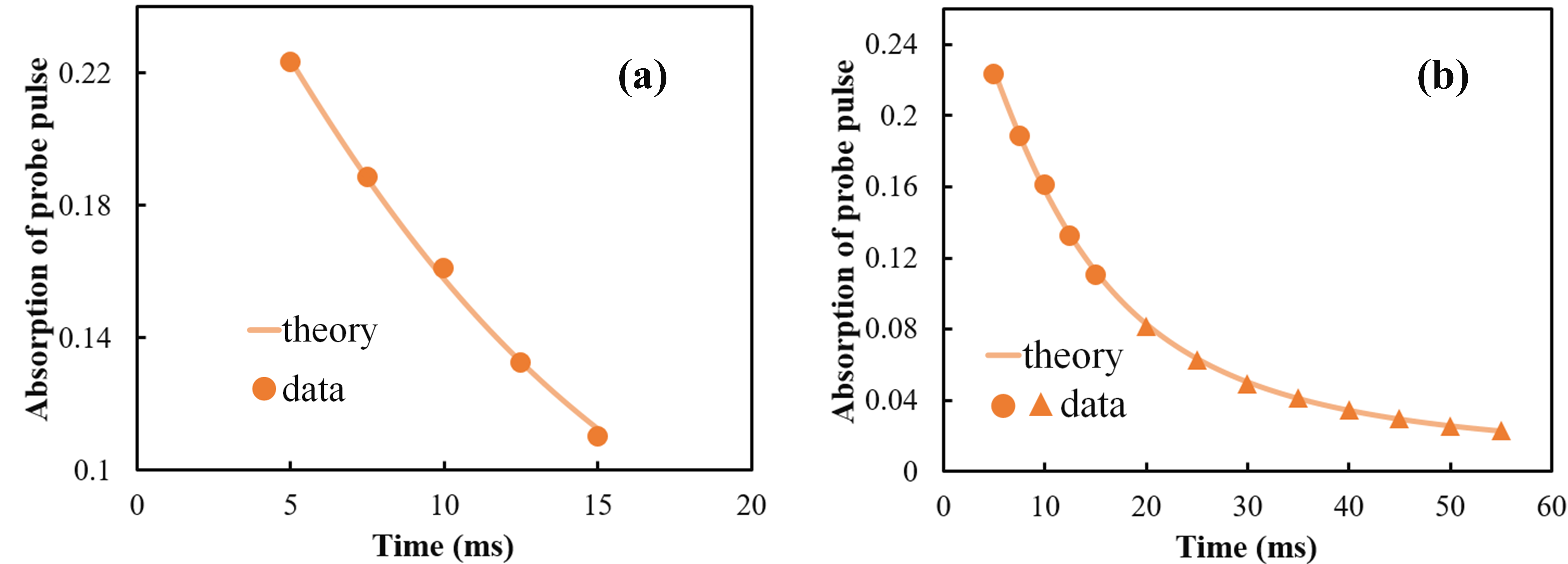}
\end{tabular}
\linespread{1}
\caption{An example of measuring temperature within a relatively short experimental time period with the nearly nondestructive method. All data points are registered from the same single experiment run. (a) Fitting to data points within the time span of the first 10 ms, which are plotted as dots. (b) Fitting to the data points of the full time span, including those of (a); these extra data points are plotted as triangles.}
\label{fig:fast_pace}
\end{figure}

Then we implement this type of nearly nondestructive method and we investigate the influence of cooling time on the temperature, as shown in Fig. \ref{fig:sample_nondestructive}. Such a test confirms that as long as the cooling time is not too short, it does not have an effect on the ultimate temperature of cold atoms in ILC. As an advantage of the nondestructive method, a complete trace of measurement can be accomplished within one experiment cycle without changing the internal state of the labeled atoms. This indicates that for practical applications, we can assess the temperature of the cold atom ensemble in the first few milliseconds or even sooner, while keeping the majority of atoms' internal states unperturbed, ready for the next phase of experiments. More specifically, an example of obtaining temperature value within 10 ms of experimental time consumption is shown in Fig. \ref{fig:fast_pace}. We anticipate that with refined techniques and better signal-to-noise ratio, the time consumption can be further reduced.

Next we investigate the influence of cooling laser power on the temperature, and the result is shown in Fig. \ref{fig:nondestructive_compare}. Meanwhile, we also include a direct comparison between the nearly nondestructive type and the basic destructive type measurements. This sample data yields identical temperature value from both cases, which verifies the feasibility of the nondestructive type method. In principle, one main limiting factor comes from the heating of the cold atoms caused by the scattering of the probe light, and this requires us to keep a relatively low power setting. For our experiment the probe light power is kept on the level of less than 1 $\mu$W.

As stated earlier, we choose the form of hollow beam in order to feed cooling laser into the cylindrical cavity for ILC. This recently proposed design is an update based on our previous studies \cite{Zhang:09_OE, PhysRevA.92.062101, Meng_2013_CPL}, with the aim of generating a more uniform intracavity diffuse optical field and enhancing the concentration of atoms along the central axis. While the performance in general receives obvious improvement, we observe that the measured temperature values seem to be consistently much lower than the Doppler temperature $T_{D}\approx145.6$ $\mu$K of $^{87}$Rb. Phenomena of sub-Doppler cooling have been extensively observed in ILC experiments of various instantiation forms, although there still exist some puzzles on the fundamental mechanisms. For instance, Ref. \cite{PhysRevA.82.033436} proposes that the underlying principle can be explained as Sisyphus cooling in a speckle laser field \cite{PhysRevA.58.3953, Grynberg_2000_EPL}.

\begin{figure}[bh]
	\centering
	\includegraphics[width=0.45\textwidth]{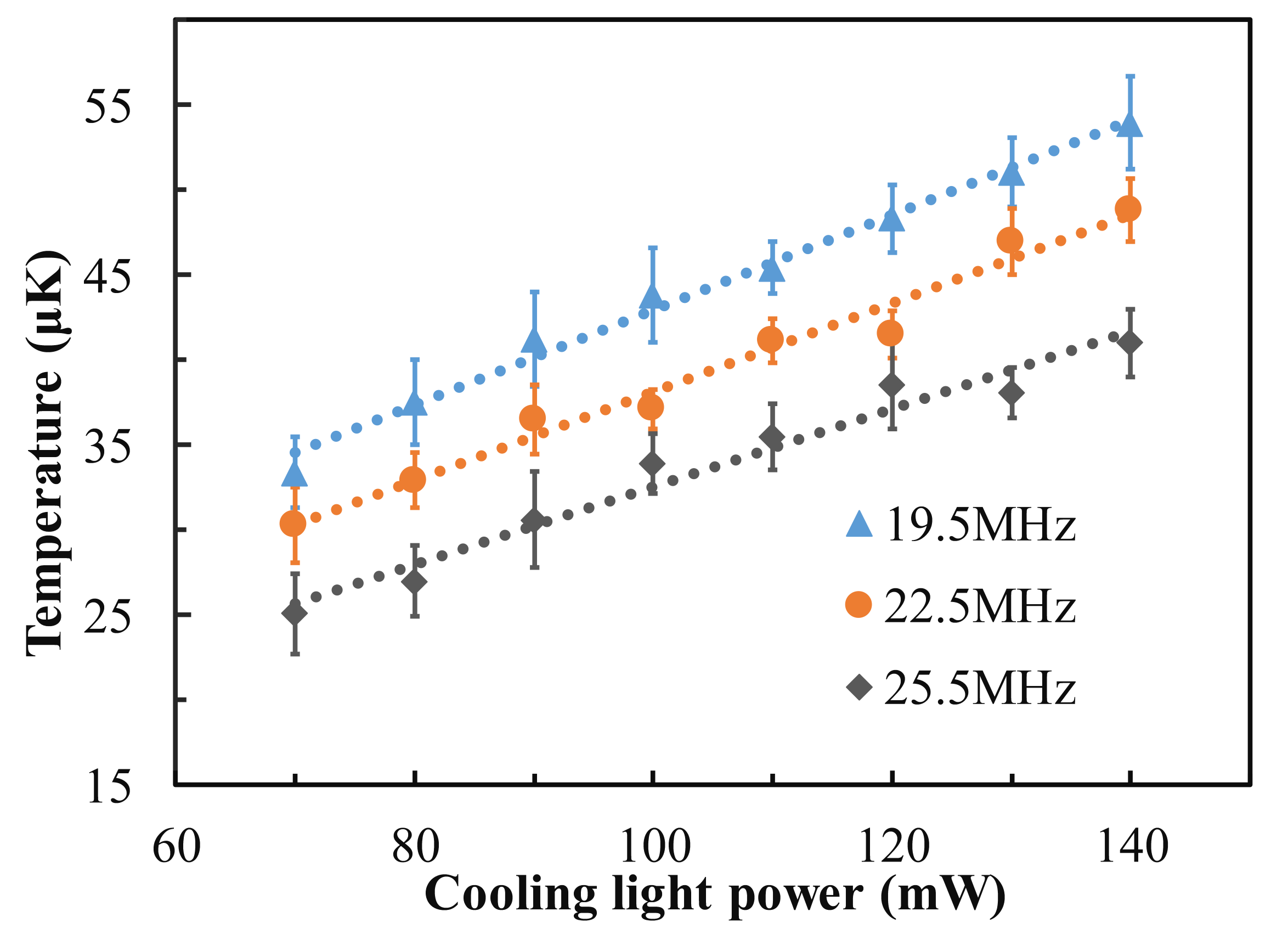}
	\caption{Relation between the cooling laser power $\mathcal{I}$ and the cold atom temperature $T$ in ILC for selected cooling laser detunings. Linear curve fitting is shown, where the intercepts on the temperature axis are around $\sim$10 $\mu$K \cite{SuppInfo}. The cooling time is set at 1000 ms for all these cases. We note that this constitutes an analogy to measurements of temperature scaling laws in optical molasses \cite{PhysRevA.57.401} and MOT \cite{Vorozcovs:05}.}
\label{fig:temperature_vs_laser_power}
\end{figure}

Following the observation of cooling laser power's influence, we carry out further investigations into the relation between temperature and cooling laser power, and a typical result is shown in Fig. \ref{fig:temperature_vs_laser_power}. It provides an insight into the physics of temperature scaling law of ILC, and in particular it hints about the embedded sub-Doppler cooling properties.

In fact, we may deduce the simplified 1D form of the sub-Doppler cooling force profile in terms of $F = -\beta_\text{sub}(\delta) v$ from elementary analysis based on Fig. \ref{fig:temperature_vs_laser_power}, with $\delta$ being the cooling laser detuning. More specifically, under the framework of Fokker-Planck equation to interpret laser cooling, the temperature limit of $k_B T=mv^2/2$ is achieved when the magnitudes of cooling rate $F \cdot v$ and the heating rate $4\hbar \omega_r \gamma_p$ are equal, with $\omega_r$ being the recoil frequency $\omega_r = \hbar k^2/(2m)$ and $\gamma_p$ being the effective scattering rate. On basis of the apparently linear relation of Fig. \ref{fig:temperature_vs_laser_power}, we can neglect the relatively small term of intercepts and treat the relation as $T = \eta(\delta)\cdot\mathcal{I}$. By definition the saturation parameter is proportional to intensity $s_0 = \alpha \mathcal{I}$ and we have the approximate relation at low intensity: $\gamma_p \approx s_0 \gamma \big(1 + (2\delta/\gamma)^2\big)^{-1}/2$ with $\gamma$ being the natural line width. Then eventually we arrive at the following equation as the 1D description:
\begin{equation}
\label{eq:subDoppler_force_profile}
|F(\delta)| \approx  \frac{1}{2} \frac{\alpha}{k_B \eta(\delta)} \frac{\hbar^2 k^2 \gamma }{1 + (2\delta/\gamma)^2} 
\cdot|v|. 
\end{equation}
Therefore, the information of $\beta_\text{sub}$ can be extracted according to $\eta(\delta)$ from linear fittings of Fig. \ref{fig:temperature_vs_laser_power} and experimentally measured value of $\alpha$. Keeping with 1D description, the ratio $\alpha$ is on the order of $\sim 0.01$ per mW for our system. And then according to Eq. \eqref{eq:subDoppler_force_profile}, from the collected data we can estimate that $|\beta_\text{sub}/\beta_\text{OM}| \gg 1$, where $\beta_\text{OM}$ is the damping coefficient of typical Doppler cooling process in 1D optical molasses \cite{PhysRevA.96.023411}.

Among many potential applications of this method, we note two particular examples: cold atoms in space missions and ultracold atoms such as BEC. Quantum technologies based on cold atoms in space missions have become a focus in the research frontier recently \cite{nature41586Becker, nat.comm.9.2760, Tino2019}. Several characteristics of our method appear attractive for this purpose: the basic principles fit the scenario of micro-gravity, the detection process is relatively fast and the apparatus is suitable for integration into a compact system. Nondestructive thermometry of ultracold atoms, especially BEC, has constantly attracted a lot of interest. We hope that our method will be helpful in this direction as well. One of the delicate points is that the heating effect must be kept at a very low level throughout the entire process \cite{Wigley:16}. For typical alkali atoms, whilst the labeling process between the two hyperfine ground states can be driven by a Raman transition to avoid heating, the probe light has to be kept at an appropriately low power level to maintain an small number of overall scattered photons. Moreover, we will also explore potential possibilities of extending our result with respect to the emerging topic of quantum thermometry \cite{Mehboudi_2019} in future work.

In conclusion, we have designed, realized and characterized a new method to measure the temperature of a labeled segment of cold atoms, with the focus on the experimental applications in ILC. Moreover, we have further developed a nearly nondestructive form of this method by utilizing the polarization degrees of freedom to enforce a proper cycling transition. This method is straightforward to implement and can yield information of temperature within several milliseconds. It has a unique feature to address the atoms along the optical detection axis, which is of essential value to ILC. With the help of this method, we have systematically studied the influence of cooling time and cooling laser power on the outcome cold atom temperature of ILC. Meanwhile, we have implemented the recently proposed technique of hollow beam incidence for intracavity diffuse light in our specific configuration of ILC. Moreover, we have investigated the sub-Doppler cooling properties of ILC by studying the relation between temperature and cooling light power.

\begin{acknowledgements}
We gratefully acknowledge support from National Key R\&D Program of China (under contract Grant No. 2016YFA0301504) and National Natural Science Foundation of China (No. 11604353). The authors also thank Peng Xu and Tian Xia for enlightening discussions.
\end{acknowledgements}

\bibliographystyle{apsrev4-1}

\renewcommand{\baselinestretch}{1}
\normalsize

\bibliography{calligraphy_ref}
\end{document}